\documentclass[12pt,preprint]{aastex}

\shortauthors{Brogan et al.}
\shorttitle{OH (1720 MHz) Maser Search}

\begin{document}

% Here is where LaTeX definitions should go.
\newcommand\HII{H\,{\sc ii}}
\newcommand\HI{H\,{\sc i}}
\newcommand\OI{[O\,{\sc i}] 63 $\mu$m}
\newcommand\CII{[C\,{\sc ii}] 158 $\mu$m}
\newcommand\CI{[C\,{\sc i}] 370 $\mu$m}        
\newcommand\SiII{[Si\,{\sc ii}] 35 $\mu$m}
\newcommand\Hi{H110$\alpha$}
\newcommand\He{He110$\alpha$}
\newcommand\Ca{C110$\alpha$}
\newcommand\kms{km~s$^{-1}$}
\newcommand\7{$\sim 7$~km~s$^{-1}$}
\newcommand\4{$\sim 4$~km~s$^{-1}$}
\newcommand\tf{25$\arcsec$}
\newcommand\fo{40$\arcsec$}
\newcommand\ft{15$\arcsec$}
\newcommand\cmt{cm$^{-2}$}
\newcommand\cc{cm$^{-3}$}
\newcommand\CeO{C$^{18}$O}
\newcommand\tCO{$^{12}$CO}
\newcommand\thCO{$^{13}$CO}
\newcommand\htco{H$_2$CO}
\newcommand\hto{H$_2$O}
\newcommand\Blos{$B_{los}$}
\newcommand\Bth{$B_{\theta}$}
\newcommand\Bm{$B_{m}$}
\newcommand\Bvm{$\mid\vec{B}\mid$}
\newcommand\BS{$B_S$}
\newcommand\Bscrit{$B_{S,crit}$}
\newcommand\Bw{$B_W$}
\newcommand\mum{$\mu$m}
\newcommand\muG{$\mu$G}
\newcommand\mjb{mJy~beam$^{-1}$}
\newcommand\jb{Jy~beam$^{-1}$}
\newcommand\thi{$\tau_{HI}$}
\newcommand\toh{$\tau_{OH67}$}
\newcommand\tmin{$\tau_{min}$}
\newcommand\tmax{$\tau_{max}$}
\newcommand\dv{$\Delta v_{FWHM}$}
\newcommand\va{$v_A$}
\newcommand\Np{$N_p$}
\newcommand\np{$n_p$}
\newcommand\We{$\mid{\cal W}\mid$}
\newcommand\Ms{${\cal M}_S$}
\newcommand\Mw{${\cal M}_w$}
\newcommand\Te{${\cal T}$}
\newcommand\Ps{${\cal P}_s$}
\newcommand\Ra{${\cal R}$}
\newcommand\Da{${\cal D}$}
\newcommand\xb{$x_b$}
\newcommand\xbt{$x_b^2$}

\title{OH (1720 MHz) Maser Search Toward the 
Large Magellanic Cloud}

\author{C. L. Brogan\altaffilmark{1,2}, W. M. Goss\altaffilmark{1},
J. Lazendic\altaffilmark{3}, A. J. Green\altaffilmark{4}}

\altaffiltext{1}{National Radio Astronomy Observatory, P. O. Box O,
1003 Lopezville Road, Socorro, NM 87801, USA; mgoss@aoc.nrao.edu}

\altaffiltext{2}{Current address: Institute for Astronomy, 640 North
A'ohoku Place, Hilo, HI 96720, USA; cbrogan@ifa.hawaii.edu} 
  
\altaffiltext{3}{Harvard-Smithsonian
Center for Astrophysics, 60 Garden Street, Cambridge, MA 02138, USA; 
lazendic@head.cfa.harvard.edu}

\altaffiltext{4}{Astrophysics Department, School of Physics,
University of Sydney, Sydney, NSW 2006, Australia; aGreen@physics.usyd.edu.au}

\email{cbrogan@ifa.hawaii.edu}

\begin{abstract}

We have carried out a sensitive search for OH (1720 MHz) masers in the
Large Magellanic Cloud (LMC) toward five regions using the Australia
Telescope Compact Array (ATCA). Our source list includes the 30
Doradus region, and four supernova remnants (SNRs): N44, N49, N120,
and N132D. These data have a typical resolution of $\sim 8\arcsec$ and
rms noise levels of 5-10 \mjb\/. We have detected OH (1720 MHz) masers
in the NE part of 30 Doradus and toward the SNR N49.  The OH (1720
MHz) maser emission in 30 Doradus is coincident with a cluster of
young stars known as ``Knot 1'', and is almost certainly of the star
formation variety. Our spectral resolution (0.68 \kms\/) is
insufficient to detect the Zeeman effect from the strongest ($\sim
320$ \mjb\/) of the 30 Doradus OH (1720 MHz) masers, leading to an
upper limit to the field strength of 6 mG.  The weak OH (1720 MHz)
maser emission (35 \mjb\/) detected toward the LMC SNR N49 is located
just west of a previously identified CO cloud, and are indicative of
an interaction between the SNR and the molecular cloud. Although the
statistics are low, the detection rate seems consistent with that seen
for Galactic star forming region and SNR type OH (1720 MHz) masers --
both of which are low.

\end{abstract}


\section{INTRODUCTION}

In recent years, it has become clear that there are at least two major
subclasses of Galactic OH (1720 MHz) masers. One is associated with
young massive star forming regions and are almost always accompanied
by mainline OH maser emission at 1665 and 1667 MHz (see, however,
\S4.1). These masers are thought to be radiatively pumped in dense gas
with $n\sim 10^7$ \cc\/, low kinetic temperatures of $\sim 30$ K, can
be quite strong $\sim 100$ Jy, and are often variable 
\citep[see for example][]{Cragg02, Caswell04}.  We will call masers
of this subclass SFR (star forming region) OH (1720 MHz) masers. In a
recent survey of 200 known mainline maser locations in the southern
sky, \citet{Caswell04} detected 34 SFR OH (1720 MHz) masers (i.e. in
$17\%$ of the sample). Caswell also finds that SFR 1720 MHz masers
occur preferentially where OH 6035 MHz masers are also found.
\citet{Szymczak04} report the results of a blind survey for
1612, 1665, 1667, and 1720 MHz masers at 100 sites of SFRs with known
6668 MHz CH$_3$OH maser emission. These authors found 55 mainline OH
maser sources and only 6 OH (1720 MHz) maser sources (all with
accompanying mainline emission) toward the methanol maser sites.  Thus
SFR OH (1720 MHz) maser emission is rare compared to mainline OH and
methanol masers.
 
The second kind of OH (1720 MHz) maser is associated with supernova
remnants (SNRs) that are interacting with nearby molecular clouds.
This subclass is not typically accompanied by mainline masers, tend to
have significantly lower brightness temperatures than their SFR
cousins, and have correspondingly large maser spot sizes \citep[see for
example][]{Frail96, Claussen99, Claussen02,
Hoffman03}.  However, a recent OH study of the SNR
G349.7+0.3 by Lazendic et al., in prep. has detected a weak OH (1665
MHz) maser in the vicinity of the OH (1720 MHz) masers. Because this
remnant is quite distant ($\sim 22$ kpc) the relationship of the two
maser types is uncertain.  SNR OH (1720 MHz) masers have been found
toward 19 SNRs, or 10\% of the known SNRs in our Galaxy \citep{Green97, 
Frail96, Koralesky98}.  In addition, a
number of masers have also been detected toward the circumnuclear disk
(CND) at the Galactic center which have properties similar to the SNR
class \citep{Yusef-Zadeh99}. Maser theory suggests that isolated
SNR OH (1720 MHz) masers can be pumped efficiently by collisions
behind C-type (non-dissociating) shocks when an SNR encounters a
nearby molecular cloud. The masers can only be efficiently pumped when
the post-shock molecular gas has densities of $\sim 10^5$ \cc\/ and
temperatures between 50 K $\lesssim T \lesssim$ 125 K 
\citep[e.g.][]{Lockett99, Wardle99}.  Indeed, CO molecular line observations 
of SNR OH (1720 MHz) masing regions \citep{Frail98} indicate
that the gas properties are in agreement with these theoretical
expectations.  Hence, observations of this OH maser line can serve as
a powerful probe of SNR/molecular cloud interactions.

From these descriptions it is clear that both SFR and SNR type OH
(1720) MHz masers are useful probes of the physical conditions present
at the locations in which they are found.  Another advantage of
observing the OH (1720 MHz) maser line is that it provides a way to
measure the strength of the magnetic field via Zeeman splitting of the
circularly polarized emission from the maser.  The recent surveys of
\citet{Caswell04} and \citet{Szymczak04} suggest that Galactic SFR
OH (1720 MHz) masers have magnetic field strengths in the range 1-16
mG.  Zeeman observations toward SNR OH (1720 MHz) masers in our Galaxy
have resulted in magnetic field detections between 0.2 and 4 mG 
\citep[see][and references therein]{Brogan00}.  

The Large Magellanic Cloud (LMC) provides a unique opportunity to
study the masers of a galaxy other than our own.  One of the
advantages of studying sources in the LMC is that their distances are
known to be near the LMC's distance of $\sim 50$ kpc since this galaxy
is viewed almost face-on \citep{Feast91}.  Moreover, the relatively close
distance, ensures that we can observe masers with typical Galactic
luminosities, rather than so called megamasers, which seem to require
very special physical conditions. For example, some of the strong
Galactic SFR type masers would have flux densities of $\sim 4$ \jb\/ at
the distance of the LMC \citep{Gaume87}, while the strongest
Galactic SNR type masers would have peak flux densities of $\sim 200$
\mjb\/ \citep{Claussen97, Yusef-Zadeh99}.

In this paper we present the results of a limited OH (1720 MHz) maser
search toward several LMC SNRs and the 30 Doradus region using the
Australia Telescope Compact Array (ATCA).  Although we have
concentrated on SNRs, a number of LMC \HII\/ regions (in close
proximity to the target SNRs) are also included in our survey. The
detection of OH (1720 MHz) masers in LMC SNRs and SFRs will serve
several purposes: (1) it will help to identify such interactions for
future molecular investigation; (2) if strong masers are observed we
will have the rare chance to detect extragalactic magnetic field
strengths directly; and (3) we will gain further insight into the
production of masers in a low metallicity environment; previous
observations of other masing lines in the LMC have yielded unusually
low detection rates \citep[see for example][]{ Scalise82, Beasley96}.

\subsection{SNR candidates}

Some $\sim 40$ LMC SNRs are currently cataloged \citep[see][]{Williams99}.
However, given the long observing times needed to detect
relatively weak OH (1720 MHz) masers, we chose SNRs with the greatest
likelihood of success for our initial survey.  Two of our candidates
are the LMC SNRs N49 (0525-66.1)  and N132D which were shown by 
\citet{Banas97} to be spatially and kinematically associated with molecular
clouds from SEST observations of CO($2-1$).  These two remnants also
show X-ray and radio limb-brightened morphologies which are thought to
arise when an SNR interacts with dense molecular material.  \citet{Dickel98} 
have recently used the ATCA to study the radio emission
from five mature SNRs in the LMC and two of these: N120 (0519-69.7),
and N44:shell 3 (0523-679) are associated with \HII\/ region
complexes, and show morphologies similar to N49.  Therefore, these
four remnants represent an ideal starting place to look for SNR OH (1720
MHz) masers in the LMC.

In addition, a brief unpublished OH (1720 MHz) maser search of the 30
Doradus region by D.  A.  Roberts (2000 private communication) with
the ATCA in 1997 has already resulted in the detection of a $\sim 380$
mJy OH (1720 MHz) maser toward the 30 Dor region.  Thus we have 
included this region in our current survey.

\section{OBSERVATIONS}

Table 1 gives the observing parameters of our 1720 MHz Australia
Telescope Compact Array observations taken between Jan.  27 and Feb.
2, 2001.  These data were taken using the 6C 6 km baseline
configuration, and a 4 MHz bandwidth.  All of the data were reduced
and imaged using the MIRIAD software package. Flux and bandpass
calibration were carried out using a single long observation of
1934-638 at the beginning of each run, while phase calibration
information was derived from more frequent observations of 0407-658.
For N44, N49, N120, and N132D we recorded two orthogonal linear
polarizations, while for 30 Dor we also recorded the
cross-correlations allowing us to image this region in RCP (right
circular polarization) and LCP (left circular polarization), as well as
Stokes I. As a result the spectral resolution of the 30 Dor data is a
factor of two worse than for the other sources. The phase calibrator
0407-658 was also used to determine the polarization leakage terms for
the 30 Doradus data.

For the four SNR fields, an average of the inner $75\%$ of the band was used
to remove the continuum emission in the $UV$ plane. For 30 Doradus the
channels containing the maser emission near the center of the band
were not used in the continuum subtraction. A continuum subtracted
line cube was then created covering the FWHM primary beam ($\sim
30\arcmin$), and using natural weighting in order to achieve the
maximum possible sensitivity. The resulting single pixel spectral rms
noise level for each source is also shown in Table 1. Each cube was
then searched for maser emission; our detections are described in \S
3. At the distance of the LMC (50 kpc) $\sim 1$ pc=$4\arcsec$. 
All velocities are presented in the LSR reference frame.

In addition to the 1720 MHz data presented here, we also reduced
archival ATCA data from May 5, 1997 at 1665 and 1667 MHz in a manner
similar to that described above in order to determine whether any
mainline masers are present toward the 30 Doradus region, and the SNRs
N49 and N44.  Approximately 2.6 hours were spent on each of these
sources at both 1665 and 1667 MHz. No mainline masers were detected
down to an rms noise level of $\sim 20$ \mjb\/ \citep[also see][who 
obtained an rms noise of $\sim 10$ \mjb\/]{Brooks97}.

\section{RESULTS}

OH (1720 MHz) masers were detected toward the NE part of 30 Doradus
and toward the SNR N49. At the distance of the LMC, thermal emission
at 1720 MHz would be undetectable, if not simply resolved out.  
No maser detections
were made toward N44, N120 (including the associated \HII\/ regions),
or N132D. The spectral line $1\sigma$ rms noise levels achieved toward
each source are listed in Table 1. Details of the OH maser detections
are described below.

\subsection{OH (1720 MHz) Masers in 30~Doradus}

In total, three distinct OH (1720 MHz) maser spots are detected toward
30 Doradus near $05{\rm^h}38{\rm^m}45.0{\rm^s}$,
${-69}\arcdeg05\arcmin07.5\arcsec$.  Figure 1 shows a Stokes I= RCP+LCP image
of the strongest 30 Doradus OH (1720 MHz) maser channel at
$v_{lsr}=243.0$ \kms\/. In addition to the strongest spatially
unresolved maser (30Dor$\_$OH(1)) in Figure 1, with a peak flux
density of I$\sim 318$ \mjb\/, there is a second weaker maser
spot $\sim 10\arcsec$ to the SE (30Dor$\_$OH(2)) with a peak Stokes I
flux density of $\sim 28$ \mjb\/ ($\sim 6\sigma$). Figure 2 shows
the RCP and LCP profiles toward the peak 30Dor$\_$OH(1) maser
position.  The third 30 Doradus OH maser (30Dor$\_$OH(3)) is apparent
in Fig. 2 at a velocity of 245.1 \kms\/ and peak LCP flux density of
$\sim 25$ \mjb\/ ($\sim 6\sigma$). The 30Dor$\_$OH(3) maser is located
within $\sim 1.5\arcsec$ of 30Dor$\_$OH(1). The 30Dor$\_$OH(1) and
30Dor$\_$OH(2) masers are predominantly right circularly polarized
while 30Dor$\_$OH(3) is left circularly polarized (see Fig. 2). No
linear polarization was detected down to the rms noise level of 5
\mjb\/.

These results are in good agreement with the previous epoch of
unpublished ATCA 1720 MHz data from 1997 toward 30 Doradus
(D. Roberts, 2000 private communication). The 1997 data have a peak Stokes
I flux density of 380 \mjb\/, rms noise $\sim 12$ \mjb\/, spatial
resolution of $6\arcsec$ and a velocity resolution of 0.4 \kms\/. The
apparent decrease in peak flux density of the current data is
consistent with the strongest 30 Dor maser being spectrally unresolved
with 0.82 \kms\/ velocity resolution. Therefore, the maser does not
appear to be significantly variable on a four year timescale.

An attempt was also made to measure the magnetic field strength of the
30 Doradus OH (1720 MHz) maser emission using the Zeeman
effect. Figure 2 shows the observed Stokes RCP and LCP emission
profiles.  At the current velocity separation 0.68 \kms\/ (0.82 \kms\/
velocity resolution), no frequency shift is apparent between the RCP
and LCP profiles. A velocity shift less than 0.68 \kms\/ implies that
the magnetic field strength is $\lesssim 6$ mG (using a Zeeman
coefficient of 0.114 \kms\/mG$^{-1}$). This upper limit is in good
agreement with the field strengths found for Galactic
OH (1720 MHz) masers: 1 - 16 mG (see Caswell 2004). Unfortunately, it is
not possible to observe with higher spectral resolution while
recording full polarization data on the ATCA.

The location of the OH (1720 MHz) masers is indicated on a composite
image of the ATCA 6 cm continuum with $Chandra$ X-ray contours
superposed \citep[from][]{Lazendic03} in Figure 3. The relationship
of the masers to other sources in 30 Doradus is discussed in detail in \S
4.1.

\subsection{OH (1720 MHz) Masers in N49}

Two OH (1720 MHz) maser spots are detected just west of the center of
N49, both with a center velocity of 278.2 \kms\/. The stronger of the
two masers has a peak flux density of $\sim 35$ \mjb\/ ($6\sigma$). An
image of the N49 1720 MHz continuum with OH (1720 MHz) maser emission
contours superposed is shown in Figure 4, and a Stokes I profile
toward the peak maser emission is shown in Figure 5.  In order to
verify our detection we split the N49 data into two approximately
equal halves, each containing $\sim 7.5$ hours of data and spanning
$\sim 12$ hours of $U-V$ coverage, and imaged them separately.  The
brighter maser spot is weakly detected in the two independent
images. Comparisons of the maser data with previous observations
of N49 are presented in \S 4.2. 

The SNR N49B (0525-660) is also located within the primary beam of our 
ATCA data. No masers were detected toward this source.

\section{DISCUSSION}

\subsection{Nature of the 30 Doradus OH (1720 MHz) Masers}

The OH (1720 MHz) masers detected toward 30 Doradus are likely of the
star-formation variety. The masers are located $\sim 2\arcsec$ NW of
an optical and radio peak called ``Knot 1'' by \citet[][see Fig. 
3]{Walborn87}. Recent high resolution HST optical and ground
based near infrared observations of this region show numerous massive
young stars in the vicinity of ``knot 1'', many of them still deeply
embedded \citep{Rubio98}.  Furthermore, the 30 Doradus OH (1720
MHz) masers appear to lie on the southern edge of a clump of molecular
H$_2$ gas observed by \citet{Rubio98}. The masers are also
coincident with the intersection of two large molecular clumps (in
projection) traced by CO(1--0) observed by \citet{Johansson98}
using the SEST telescope.  One, lying just NE of the maser location is
has $v_{lsr}\sim 249$ \kms\/ and $\Delta v\sim 10$ \kms\/ (cloud
30Dor-10) while the other lies just to the SW and has $v_{lsr}\sim
246.5$ \kms\/ and $\Delta v\sim 3$ \kms\/ (cloud 30Dor-12). These
velocities are in fair agreement with those of the OH (1720 MHz)
masers (243.0 \kms\/).

A recent search of the 30 Dorodus region for SNRs by \citet{Lazendic03} 
using radio, optical, and X-ray data did not yield any
detections near the maser locations. However, the masers are located
fairly close (in projection) $\sim 26\arcsec$ or 6.5 pc to a
Wolf-Rayet cluster of stars called R140 \citep[see][also see Fig. 
3]{Portegies02}. Recent $Chandra$ observations by \citet{Portegies02} 
have revealed that this cluster is very bright in
X-rays, and it is possible that the X-rays in combination with a shock
driven by the Wolf-Rayet winds could effect the same type of
conditions present in a SNR shock \citep{Wardle99}. It is unclear how
far away the WR cluster could exert such an influence. Indeed, from
Figure 3 it is clear that the X-rays fill a bubble devoid of radio
continuum emission (H$\alpha$ emission looks much the same) to the NW
of R140.  However, the edge of the bubble traced by X-rays does not
appear to reach as far east as the maser location.

Two further arguments against an SNR interpretation for the strongest 30
Doradus OH (1720 MHz) maser are its high luminosity 1150 Jy kpc$^2$
(assuming a 50 kpc distance to the LMC) and strong right circular
polarization (compared to left). The strongest Galactic SNR type OH
(1720 MHz) masers (those in W28 and SgrA) have luminosities of $\sim
500$ Jy kpc$^2$ \citep{Claussen97, Yusef-Zadeh99}.
Conversely, while not common, a few Galactic SFR OH (1720 MHz) masers
have luminosities in excess of 3,000 Jy kpc$^2$ \citep[in W49A and W51 for
example,][]{Gaume87}. With only one exception, SNR OH (1720
MHz) masers typically show equal amounts of right and left circularly
polarized emission \citep[see for example][]{Brogan00}. Even the
exception, SNR IC443 only has an excess right circular polarization of
$10\%$ \citep{Hoffman03}. In contrast the 30~Doradus OH (1720
MHz) maser has more than 2.5 times the flux density in RCP as LCP (see
Fig. 2). Preferentially stronger masing in one circular polarization
compared to the other is thought to originate from velocity gradients,
and is common in SFR type OH masers \citep[see for example][]{ 
Caswell04, Gaume87}.

It is surprising, however, that no mainline OH maser emission down to
a $5\sigma$ rms noise level of $\sim 50$ \mjb\/ ($\sim 9$ times weaker
than the 1720 MHz RCP peak flux density, Fig. 2) has been detected
\citep[see \S 2;][]{Brooks97}.  It is uncommon to find Galactic
SFR OH (1720 MHz) masers that are not also accompanied by mainline OH
masers at 1665 and 1667 MHz. Indeed, the mainline masers are often
stronger than those at 1720 MHz.  For example, of the 34 SFR OH (1720
MHz) masers detected by \citet{Caswell04} and \citet{Szymczak04}
with accompanying mainline OH maser emission, less than $10\%$ are
significantly (more than three times) stronger than the coincident
mainline masers.  However, \citep{Caswell04}, has also serendipitously
detected six isolated SFR (1720 MHz) masers \citep[also see Lazendic et al.
in prep. for CTB~33 and][for W3]{Koralesky98}.  That is, the
masers were detected within the primary beam of the observations, but
well separated from the mainline target locations. To date, no truly
blind Galactic surveys for OH (1720 MHz) masers have been carried out
so it is unclear to what extent isolated SFR OH (1720 MHz) masers
exist even in our own Galaxy. However, due to the large primary beam
of the ATCA at 18 cm ($\sim 30\arcmin$), the Caswell 1720 MHz survey
of 200 southern mainline maser locations, covers an area of
50-deg$^2$. Thus it is clear that such masers are rare. 

Despite a number of recent surveys, no H$_2$O, or CH$_3$OH masers have
been reported in the literature near this location either
\citep{Beasley96, Lazendic02}. The only
other known maser in the 30~Doradus region is an H$_2$O maser that is
located $\sim 35\arcsec$ NE of the OH (1720 MHz) maser. Its position
from \citet{Lazendic02} is also plotted on Figure 3 for reference.

To summarize, most of the evidence: coincidence with a region of high
mass star formation, high maser luminosity, and preferentially right
circularly polarized emission suggests a SFR origin for the 30~Doradus
OH (1720 MHz) maser. The absence of mainline OH masers at this
location is surprising but not exceptional. Unfortunately, \citet{Caswell04} 
also finds that unlike typical SFR OH (1720 MHz) masers, the
few known isolated SFR OH (1720 MHz) masers also lack methanol and
excited state OH masers. Thus further observation is unlikely to clarify
this issue.

\subsection{Nature of the N49 OH (1720 MHz) Masers}

Accounting for distance, the N49 OH (1720 MHz) maser flux density
$\sim 35$ \mjb\/ is within the range observed for Galactic SNR OH
(1720 MHz) masers \citep{Claussen97, Brogan00}.  As
described in \S 1, the presence of an isolated OH (1720 MHz) maser
toward the SNR N49 is indicative of an interaction between the remnant
and a molecular cloud. Several recent observations have confirmed the
presence of molecular gas in the vicinity of N49. \citet{Mizuno01}
have observed this region in CO(1--0) with $2\farcm6$ resolution using
the NANTEN telescope. N49 lies within the northwestern edge of their
CO cloud M5263-6606 ($05{\rm^h}26{\rm^m}19.9{\rm^s}$,
${-66}\arcdeg03\arcmin33\arcsec$ J2000), which has a center velocity
of 285.5 \kms\/ and $\Delta v=6.8$ \kms\/ \citep[also see CO cloud no. 23 
in][]{
Cohen88}.  Recall that the maser position is
$05{\rm^h}25{\rm^m}56.5{\rm^s}$, ${-66}\arcdeg05\arcmin01\farcs5$
(J2000) and the center velocity is 278.2 \kms\/ (Fig. 5).  \citet{Banas97} 
observed CO (2-1) emission toward the eastern and SE
regions of N49 (coincident with the continuum peak) in the velocity
range 281 to 291 \kms\/ using the SEST telescope and $23\arcsec$
resolution. The CO velocity peak, 286 \kms\/ is somewhat higher than
that of the OH maser at 278.2 \kms\/, but is still in reasonable
agreement. Especially when one considers that Banas et al.  used the
velocity range 263-280 \kms\/ for baseline subtraction, possibly
affecting the apparent extent of the CO line.  Curiously however, the
CO(2--1) emission only barely extends far enough west to encompass the
maser emission.  Indeed the CO emission is strongest at the bright SE
boundary of the SNR \citep{Banas97}. Unfortunately, Banas et
al. did not observe exactly at, or west of the maser locations.  It
would be interesting to obtain higher sensitivity, higher $J$ CO
observations at the location of the maser emission to look for
evidence of shocked thermal molecular gas.

Further evidence for an interaction comes from the presence of faint
diffuse infrared emission from the SE part of the remnant. Figure 6
shows a $J-$band (1.25 \mum\/) 2MASS image superposed with the 1720
MHz continuum and maser contours. The diffuse $J-$band near-IR
emission originates from warm dust, and possibly hydrogen
recombination lines presumably heated and ionized by the passage of
the shock \citep[see for example][for SNR IC443]{Rho01}. From the
near-IR morphology, together with that of the CO emission, we might
reasonably expect that any maser emission would be in the southeastern
region. However, \citet{Lockett99} find that the OH (1720 MHz)
maser collisional pump becomes less efficient for higher dust optical
depths and temperatures ($T_d\gtrsim 100$ K). In this case, it is
then, not surprising to see a lack of masers toward the regions of
near-IR dust emission. Additionally, N49 has the highest optical
surface brightness of any SNR in the LMC \citep{Vancura92}. Like
the radio continuum, the optical emission is strongest toward the
southeastern regions of the SNR, although it is more extended than the
near-IR emission shown in Fig. 6.  From optical spectroscopy \citep{Vancura92}
find that about 1/3 of the optical emission comes from
shocks with speeds $\gtrsim 200$ \kms\/, while another 1/2 comes from
shocks with speeds of $\sim 100$ \kms\/. Such high shock speeds are
also not conducive to SNR OH (1720 MHz) maser emission, since the
collisional pump requires slower C-type (non-dissociating) shocks in
order to form a sufficient column of OH.

Interaction with the molecular cloud is also evident from the
properties of the X-ray emission from N49. Figures 7a,b show the hard
and soft X-ray emission from N49 observed by $Chandra$ \citep{Park03}. 
\citet{Park03} suggest that the soft and hard X-ray
emission observed toward the eastern part of the SNR originates from a
decelerated shock transmitted through the cloud, and from a shock
reflected off the cloud back into the SNR interior, respectively.  In
Figure 7a,b, the X-ray emission from SGR 0526-66 is apparent
(especially in hard X-rays) towards the NE, in addition to more
extended emission from the SNR, concentrated to the SE. The SGR is
thought to be the neutron stellar remnant of the N49 progenitor star.
There is no obvious correlation of X-ray emission with the maser
locations, although there is some indication of an enhancement in the
hard X-rays (Fig. 7a). Most interestingly, there is little evidence
that the soft X-rays are absorbed toward the SE part of the remnant
where the CO, IR, and radio continuum emission peak. This may be an
indication that the bulk of the CO cloud lies behind the SNR.

In many respects, N49 closely resembles the Galactic OH (1720 MHz)
maser SNR 3C391. For this SNR, the brightest CO (2-1) emission is also
coincident with the brightest radio continuum emission, but one of the
3C391 masers lies well away from the radio continuum and CO emission
peaks \citep{Frail96, Reach99}. In the case of 3C391, it 
is clear that the bulk of the CO cloud lies in front of the SNR, because the 
soft X-rays are heavily absorbed toward the peak CO and radio continuum 
regions \citep{Chen01}. This displacement of the
maser emission from what seem to be the most obvious locations,
i.e. places where there is a very clear indication of an interaction,
is likely due to a combination of the physical conditions and the 
need for velocity coherence along the line of sight. Thus although 
an SNR may be interacting with a molecular cloud over a large surface 
area, only in those places where the temperature and density fall within 
the range needed for the collisional pump can masers form. Additionally, 
since masers are beamed, we can only observe masers where the 
shock front happens to be moving perpendicular to our line of sight, hence 
providing the requisite velocity coherence and beaming in our direction.  

\subsection{Detection Statistics}

Although it is known that Galactic SFR OH (1720 MHz) masers are
relatively rare, as mentioned previously, no truly blind survey of
star forming regions has been carried out. When sources are selected
based on previous detections of either mainline OH or methanol masers,
OH (1720 MHz) masers are detected 10-20\% of the time \citep{Szymczak04, 
 Caswell04}. However, the recent serendipitous
detection of six isolated SFR OH (1720 MHz) masers by \citet{Caswell04},
suggests that such masers could be more numerous than previously
suspected. In any case, based on the currently known data, it is not
surprising that we only detect one SFR variety OH (1720 MHz) maser in
30~Doradus. What is actually more surprising is the lack of mainline
OH, H$_2$O and CH$_3$OH masers in the LMC \citep[see][]{Scalise82,
Beasley96}.

Of the 19 Galactic SNRs with positive OH (1720 MHz) maser detections
($\sim 10\%$ of known Galactic SNRs), six ($30\%$) would have at {\em
least} one maser with a flux density $\gtrsim 30$ mJy at the distance
of the LMC \citep{Frail96, Green97, Claussen97, Koralesky98}.  
A number of OH (1720 MHz) masers
have also been observed toward the circumnuclear disk (CND) around the
Galactic center which have properties similar to those found in SNRs,
and would also be detectable at the distance of the LMC 
\citep{Yusef-Zadeh99}. Based on these statistics it is not surprising that we
detect SNR OH (1720 MHz) maser emission from one of the seven SNRs
surveyed.  \citep[including the two 30~Doradus SNR candidates 
from][]{Lazendic03}.

\section{SUMMARY AND CONCLUSIONS}

In an ATCA search for OH (1720 MHz) masers toward 30~Doradus
and four LMC supernova remnants, we have discovered two new maser
regions. To our knowledge these are the first detections of OH (1720
MHz) masers in the LMC. One region of maser emission is coincident
with the young star cluster in 30~Doradus known as ``Knot 1'', and is
almost certainly of the star formation variety. Our spectral
resolution (0.68 \kms\/) is insufficient to detect the Zeeman effect
from the strongest of the 30~Doradus OH (1720 MHz) OH masers, leading
to an upper limit to the field strength of 6 mG, in good agreement
with the fields found in similar Galactic masers. The 30
Doradus OH (1720 MHz) masers are somewhat unusual for SFR masers, since no
mainline masers are detected in their vicinity. However, a few
isolated SFR OH (1720 MHz) masers have also recently been detected in our
Galaxy \citep{Caswell04}. The other region of OH (1720 MHz) maser
emission was detected toward the LMC SNR N49. The masers are located 
just west of a CO molecular cloud detected by \citet{Banas97}, and
are indicative of an interaction between the SNR and the molecular
cloud. Although the statistics are low, the detection rate seems
consistent with that seen for Galactic SFR and SNR type OH (1720 MHz)
masers -- both of which are low.

\acknowledgments

We are very grateful to D. Roberts for his help with this
project and alerting us to the presence of the 30 Doradus maser.  We 
thank J. Caswell for his valuable insight and comments on the manuscript. The
authors also thank S. Park for providing us with his N49 $Chandra$
X-ray data in digital form, and M. Wardle for valuable discussions on
whether R140 could be responsible for exciting the 1720 MHz masers in
30~Doradus. We have made use of the 2MASS archive which is a joint
project of the University of Massachusetts and the Infrared Processing
and Analysis Center/California Institute of Technology, funded by the
National Aeronautics and Space Administration and the National Science
Foundation.

\newpage

\begin{figure}[h!]
\epsscale{0.4}
\plotone{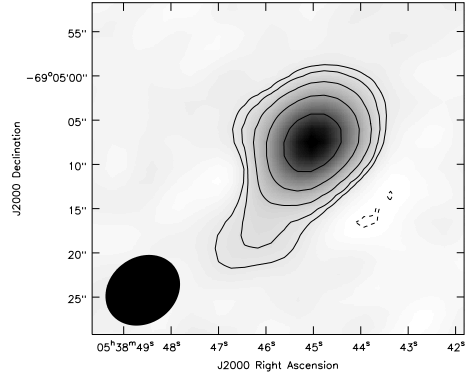}

\caption[]{ATCA Stokes I image of the strongest 30~Doradus OH (1720 MHz)
maser channel at $v_{lsr}=243.0$ \kms\/. The contour levels are at $-15, 15,
25, 50, 100$, and 200 \mjb\/. The peak flux density is 318 \mjb\/. The
resolution of this image is $8.9\arcsec\times 7.4\arcsec$
P.A.$=-54\arcdeg$. The beam size is shown on the lower left corner of
the image, and the rms noise in the image is 5 \mjb\/.}

\end{figure}

\begin{figure}[h!]
\epsscale{0.3}
\plotone{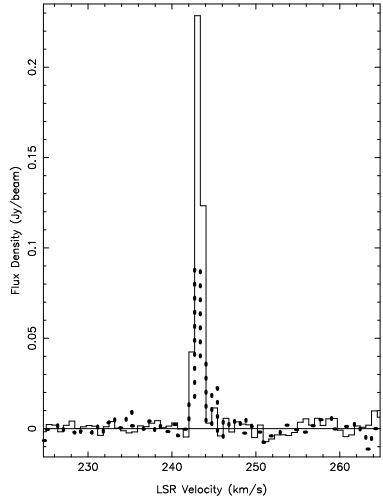}

\caption[]{ATCA Stokes RCP (solid) and LCP (dotted) line profiles of
the 30~Doradus maser at the peak position. The position of the peak is
$05{\rm^h}38{\rm^m}45{\rm^s}$, ${-69}\arcdeg05\arcmin07\arcsec$ (J2000). 
}

\end{figure}

\begin{figure}[h!]
\epsscale{0.9}
\plotone{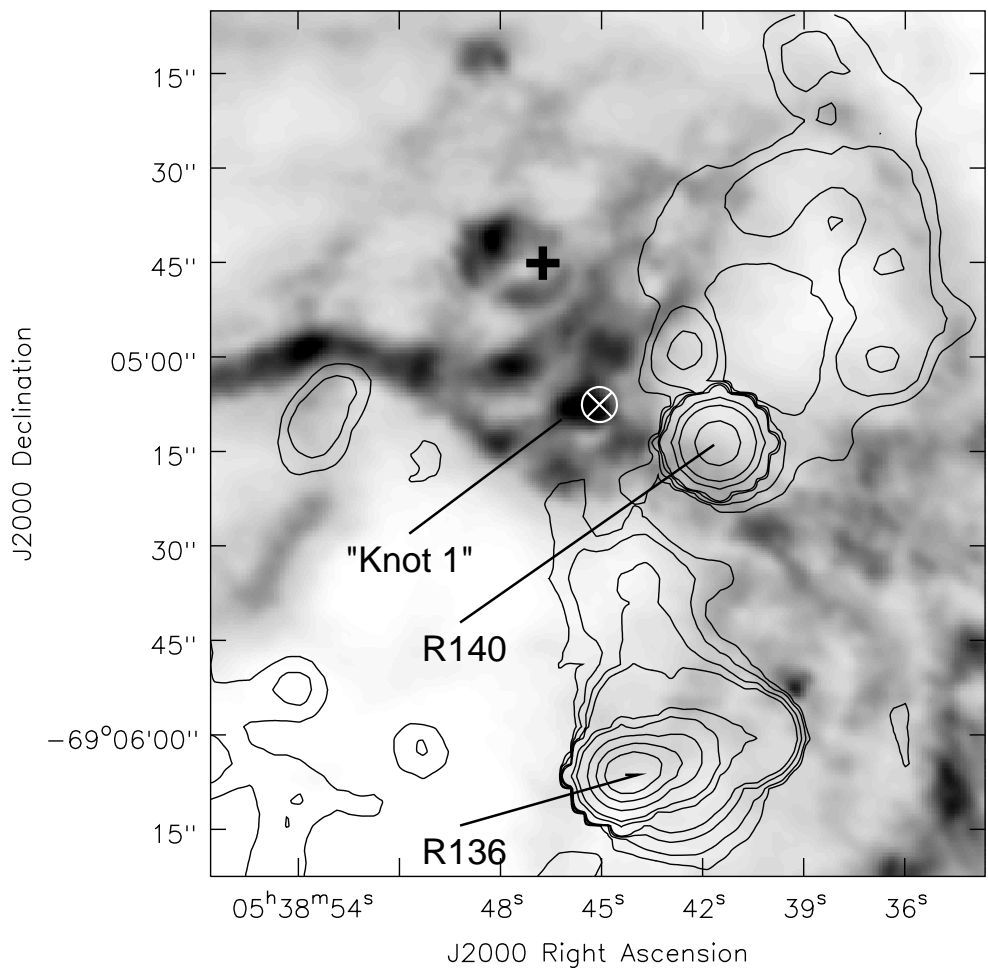}

\caption[]{Composite image of 30~Doradus showing the ATCA 6cm
continuum in greyscale and $Chandra$ X-ray contours from Lazendic et
al. (2003). The resolution of the 6 cm continuum and X-ray images are
$1\farcs8\times 1\farcs7$ and $2\arcsec$, respectively. The location
of the OH (1720 MHz) masers is indicated by the white $\bigotimes$
symbol.  The position of the only other known maser in the 30~Doradus
region: an H$_2$O maser from \citet{Lazendic02} is indicated by
the black $+$ symbol. The locations of the young star forming cluster
``Knot 1'' and the two Wolf-Rayet X-ray star clusters R140 and R136
are also indicated for reference.}

\end{figure}

\begin{figure}[h!]
\epsscale{0.6}
\plotone{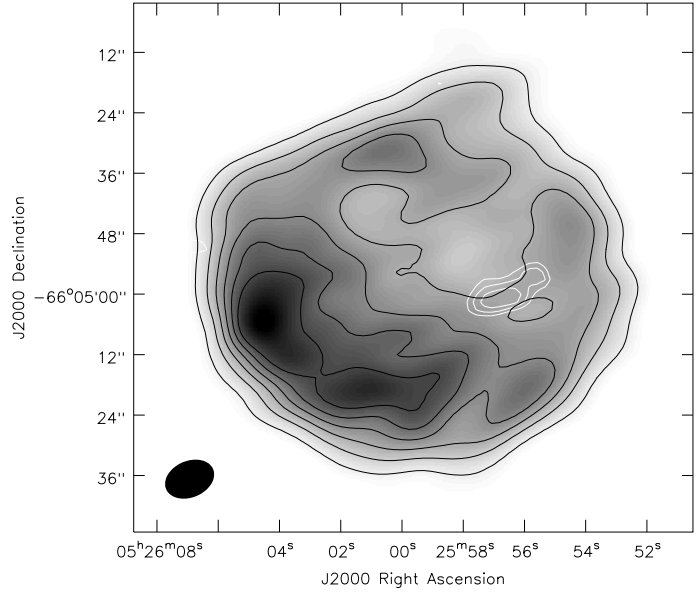}

\caption[]{Composite image of the N49 1720 MHz continuum (greyscale
and black contours) superposed with white Stokes I OH (1720 MHz) maser emission
contours from the peak channel at $v_{lsr}=278.2$ \kms\/. The resolution of the
images is $10.1\arcsec\times 7.2\arcsec$ P.A. $-67\arcdeg$ (beam is
indicated in lower left of the image). The continuum contours levels
are 5, 10, 20, 30, 40 \mjb\/ and the continuum peak flux density is 47
\mjb\/.  The maser contours levels are 20, 25, 30 \mjb\/ and the peak
maser flux density is 35 \mjb\/. The continuum rms noise is 0.6 \mjb\/, 
while the rms noise in the spectral line is 5.8 \mjb\/.}

\end{figure}

\begin{figure}[h!]
\epsscale{0.3}
\plotone{Brogan.Fig5.eps}

\caption[]{ATCA Stokes I line profile of
the N49 OH (1720 MHz) maser at the peak position. The position of the peak is
$05{\rm^h}25{\rm^m}56.5{\rm^s}$, ${-66}\arcdeg05\arcmin01.5\arcsec$ 
(J2000). }

\end{figure}

\begin{figure}[h!]
\epsscale{0.6}
\plotone{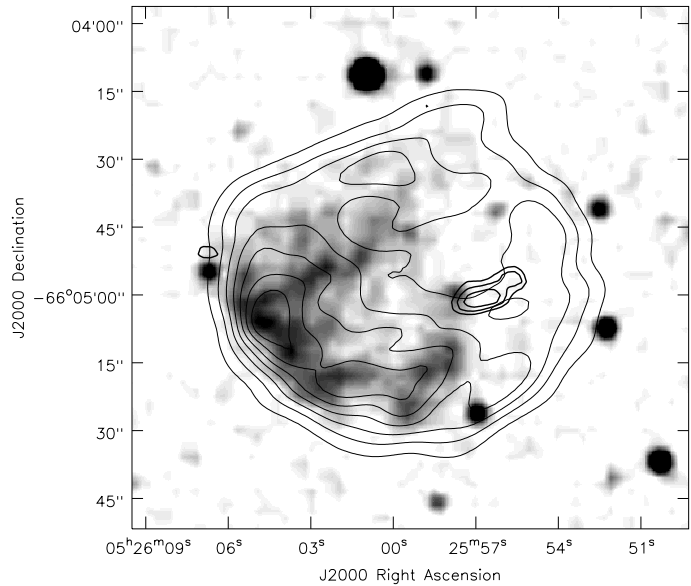}

\caption[]{2MASS 1.25 \mum\/ $J-$band image with the 1720 MHz continuum
and maser emission contours from Figure 4 superposed. }

\end{figure}

\begin{figure}[h!]
\epsscale{1.0}
\plotone{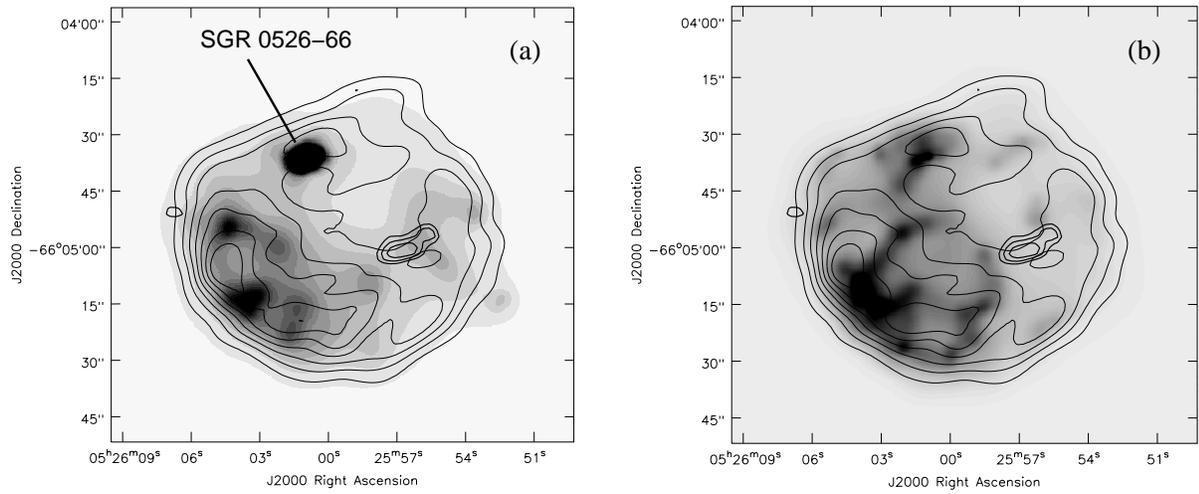}

\caption[]{ATCA 1720 MHz continuum (black) and OH (1720 MHz) maser
(white) emission contours from Figure 4 superposed on (a) $Chandra$ hard
(1.6-8.0 keV) X-ray greyscale and (b) $Chandra$ soft (0.3-0.75 keV)
X-ray greyscale from \citet{Park03}. The X-ray images have been
adaptively smoothed.}

\end{figure}

\clearpage

\newpage

\begin{deluxetable}{lccccc}
\small
\tablewidth{46pc}
\tablecaption{Observational Parameters of ATCA OH (1720 MHz) maser Observations}
\tablehead{
\colhead{Name}  & \colhead{Position$^a$} & \colhead{Integration$^b$} & \colhead{beam} & \colhead{Channel Width$^c$} & \colhead{Line RMS} \\
  & \colhead{(J2000)}  & \colhead{(hours)} & \colhead{($\arcsec\times\arcsec$ (P.A.$\arcdeg$))} & \colhead{(km s$^{-1}$)} & \colhead{(mJy beam$^{-1}$)} }     
\startdata
30~Doradus & $05{\rm^h}38{\rm^m}45{\rm^s}$, ${-69}\arcdeg$05\arcmin07\arcsec &  21.9 & $8.9\times7.4$ ($-54$) & 0.68 & 4.3\\
N49 & $05{\rm^h}26{\rm^m}00{\rm^s}$, ${-66}\arcdeg$05\arcmin00\arcsec & 15.7 & $10.1\times7.2$ ($-67$)  & 0.34 & 5.8\\
N44 & $05{\rm^h}23{\rm^m}18{\rm^s}$, ${-67}\arcdeg$56\arcmin00\arcsec  &  8.1 & $10.7\times 6.8$ ($-25$)  & 0.34  & 8.5 \\
N120 & $05{\rm^h}18{\rm^m}42{\rm^s}$, ${-69}\arcdeg$39\arcmin30\arcsec  & 9.0 & $10.7\times6.4$ ($-30$)  & 0.34  & 7.1 \\
N132D & $05{\rm^h}25{\rm^m}02{\rm^s}$, ${-69}\arcdeg$38\arcmin36\arcsec & 7.8 & $9.8\times6.8$ ($-20$)   & 0.34  & 8.3  
\label{tab1}\enddata
\tablenotetext{a} {Positions from Williams et al. (1999) except for 30~Doradus which is the position of the 
observed OH (1720 MHz) maser.}
\tablenotetext{b} {Approximate time on source.}
\tablenotetext{c} {The spectral resolution is $1.2\times$ channel width.}
\end{deluxetable}

\end{document}